\documentclass[11pt,a4paper]{article}
\usepackage{amsfonts,amssymb,latexsym, amsmath,epsfig,theorem}
\usepackage[latin1]{inputenc}
\usepackage{fullpage}
\usepackage{algorithm2e}
  \usepackage{mathptmx}
  \usepackage{textcomp}

\newcommand{\DP}[2]{ \ensuremath{ \frac{\partial #1 }{\partial #2 } } }
\newcommand{\D}[2]{ \ensuremath{ \frac{d #1 }{d #2 } }}

\newcommand{\noun}[1]{\textsc{#1}}

\newcommand\BS[1]{\mbox{\fontseries{b}\selectfont #1}}
\newcommand\plus{\BS{+}}
\newcommand\moins{\BS{\textminus}}

\newcommand\indet{\BS{?}}
\newcommand\zero{\BS{0}}

\theoremstyle{break}\newtheorem{Theorem}{Theorem}[section]
\theoremstyle{break}
\theoremstyle{break}
\theoremstyle{break}

\theoremstyle{break}\newtheorem{System}{System}
\theoremstyle{break}
\theoremstyle{break}\newtheorem{Observations}{Observations}

\DeclareMathOperator{\pred}{pred}
\DeclareMathOperator{\sgn}{sgn}

\begin{document}

\title{Complex Qualitative Models in Biology: a new approach}

\author{ P. \noun{Veber}$\ ^1$
\and M. \noun{Le Borgne}$\ ^1$ \and A. \noun{Siegel}$\ ^1$
\and S. \noun{Lagarrigue}$\ ^3$ \and
O. \noun{Radulescu}$\ ^2$ }

\date{}

\maketitle

{\footnotesize $^1$ Projet Symbiose.  Institut de Recherche en
Informatique et Syst\`emes Al\'eatoires, IRISA-CNRS 6074-Université de
Rennes 1, Campus de Beaulieu, 35042 Rennes Cedex, France} 

{\footnotesize $^2$ Institut de Recherche
Math\'ematique de Rennes, UMR-CNRS 6625, Université de Rennes 1, Campus de Beaulieu, 35042 Rennes Cedex, France}

{\footnotesize  $^3$ UMR Génétique animale, Agrocampus Rennes-INRA,  65 rue de Saint-Brieuc,  CS 84215 Rennes, France}

\smallskip
\paragraph{Abstract.} 
{\em 
We advocate the use of qualitative models in the analysis of large biological
systems. We show how qualitative
models are linked to theoretical differential models and practical
graphical models of biological networks. A new technique for analyzing qualitative models is
introduced, which is based on an efficient representation of
qualitative systems. As shown through several applications, this
representation is a relevant tool for the understanding and testing of
large and complex biological networks.
}

\section{Introduction}
\label{sec:intro}

Understanding the behavior of a biological system from the interplay
of its molecular components is a particularly difficult task. A
model-based approach proposes a framework to express some hypotheses
about a system and make some predictions out of it, in order to compare
with experimental observations. Traditional approaches (see
\cite{DJ02} for
an interesting review) include ordinary differential equations or
stochastic processes. While they are powerful tools to acquire a fine
grained knowledge of the system at hand, these frameworks need
accurate experimental data on chemical reactions kinetics, which are
scarcely available. Furthermore, they also are computationally
demanding and their practical use is restricted to a limited number of
variables.

As an answer to these issues, many approaches were proposed, that
abstract from quantitative details of the system. Among others, let us
stress the work done on gene regulation dynamics \cite{DJGHP04+}, hybrid
systems \cite{GT04} or discrete event systems \cite{CRCDF04+},
\cite{CRRT04}. The goal of such
qualitative frameworks is to enable system-level analysis of a
biological phenomenon. This appears as a relevant answer to recent
technical breakthrough in experimental biology:
\begin{itemize}
\item microarrays, mass spectrometry, protein chips currently allow to
  measure thousands of variables simultaneously,
\item obtained measurements are rather noisy, and may not be
  quantitatively reliable.
\end{itemize}

Microarrays for instance, are used for comparing the activity of genes
between two experimental settings. A microarray experiment gives
differential measure between two experimental settings. 
It delivers  informations on the relative activity of each
gene represented on the array. Despite many attempts made to
quantified the output of microarrays, the essential output of the
technique says, for example, that a gene
G is more active in situation A than in situation B.

In this paper, we use a framework developed in \cite{Biosystems} for
the comparison of two experimental conditions, in order to derive
qualitative constraints on the possible variations of the
variables. Our main contribution is the use of an efficient
representation for the set of solutions of a qualitative system.
This representation allows to solve  systems with
hundreds of variables. Moreover, this representation opens the way to finer
analysis of qualitative systems. This new approach is
illustrated by solving three important problems:
\begin{itemize}
\item checking the accordance of a qualitative system with qualitative
  experimental data.
\item minimally correcting corrupted data in discordance with a
  model
\item helping in the design of experiments
\end{itemize}

Our main focus here is to show how to use large qualitative models and
qualitative interpretations of experimental data. In this respect our
work could be used as an extension to what was proposed in \cite{GRRLH03+},
where basically the authors propose to analyze pangenomic gene
expression arrays in \emph{E.coli}, using simple qualitative rules. 

In the first section we establish links between differential,
 graphical and qualitative models.

\section{Mathematical modeling}
\label{sec:maths} In this section we show how qualitative models can
be linked to more traditional differential models. Differential
models are central to the theory of metabolic control
\cite{meta-control,meta-control2}. They also have been
applied to various aspects of gene networks dynamics.
The purpose of this section is to lay down a set of qualitative
equations describing steady states shifts of differential models.
For the sake of completeness, we rederive in a simpler case results
that have been established in greater generality in
\cite{Biosystems,Radulescu05}.

\subsection{Modeling assumptions}

Let us consider a network of interacting cellular constituents,
numbered from 1 to $n$. These constituents may be proteins, RNA
transcripts or metabolites for instance. The state vector $X$
denotes the concentration of each constituent.

\paragraph{Differential dynamics}

$X$ is assumed to evolve according to the following differential
equation:

$$ \D{X}{t}=F(X) $$

\noindent where $F$ is an (unknown) nonlinear, differentiable function. A
steady state $X_{eq}$ of the system is a solution of the algebraic
equation:

$$ F(X_{eq}) = 0.$$

Steady states are asymptotically stable if they attract all nearby
trajectories. A steady state is non-degenerated if the Jacobian
calculated in that steady state is non-vanishing.
 According to the Grobman-Hartman theorem, a sufficient condition to have
 nondegenerated asymptotically stable
 steady states is
$Re (\lambda_i) < -C, C>0, i=1,\ldots,n$, where $\lambda_i$ are the
eigenvalues of the Jacobian matrix calculated at the steady state.

\paragraph{Experiment modeling}
Typical two state experiments such as differential microarrays are
modeled as steady state shifts. We suppose that under a change of
the control parameters in the experiment, the system goes from one
non-degenerated stable steady state to another one. 
The output of the two state experiment can be expressed in terms of
concentration variations for a subset of products, between the two
states. We suppose that the signs of these variations were proven to
be statistically significant.

\paragraph{Interaction graph}
The only knowledge we require about the function $F$ concerns the
signs of the derivatives $\DP{F_i}{X_j}$.  These are interpreted as
the action of the product $j$ on the product $i$. It is an
activation if the sign is $\plus$, an inhibition if the sign is
$\moins$. A null value means no action.

An interaction graph $G(V,E)$ is  derived from the Jacobian matrix
of $F$:
\begin{itemize}
\item with nodes $V = \{1,\dots,n\}$ corresponding to products
\item and (oriented) edges $E = \{ (j,i) | \DP{F_i}{X_j} \neq 0
  \}$. Edges are labeled by $s(j,i) = \sgn(\DP{F_i}{X_j})$.
\end{itemize}

The set of predecessors of a node $i$ in $G$ is denoted $\pred(i)$. The 
interaction graph is actually built from informations gathered
in the literature. In consequence in some places it may be
incomplete (some interactions may be missing), in others it may be
redundant (some interactions may appear several times as direct and
indirect interactions). It is an important issue that neither
incompleteness nor redundancy do not introduce inconsistencies and
this will be addressed in section \ref{sec:qua_exp}.

\paragraph{Negative diagonal in the Jacobian matrix}
For any product $i$, we exclude the possibility of vanishing
diagonal elements of the Jacobian $\DP{F_i}{X_i}$. This can be
justified by taking into account  degradation and dilution (cell
growth) processes that can be represented  as negative self-loops in
the interaction graph, that is for all $i$, $(i,i) \in E$ and
$s(i,i) = \moins$.

\paragraph{Discussion} 
 In our mathematical modeling we suppose that the system
starts and ends in non-degenerated stable steady states. Of course
this is not always the case for several reasons: the waiting time to
reach steady state is too big; one can end up in a limit cycle and
oscillate instead of reaching a steady state.
All these possibilities should
be considered with caution.  Actually this hypothesis 
might be difficult to check from the two states only. Complementary
strategies such as time series analysis 
could be employed in order to assess the possibility of limit cycle
oscillations.

Positive self-regulation is also  possible but introduces
a supplementary complication.
In this case
for certain values of the  concentrations
degradation exactly compensates the positive self-regulation and
the diagonal elements of the Jacobian vanish (this is
a consequence of the intermediate value theorem).
We can avoid dealing with this situation by
considering that the positive self-regulation does not act directly
and that it involves intermediate species.
This is a realistic assumption because a  molecule never really acts directly
on itself (transcripts can be auto-regulated but only via protein
products).
Thus, all nodes can keep their negative self-loops and all diagonal
elements of the Jacobian can be considered to be non-vanishing.
Although the positive regulation may imply vanishing higher order
minors of the Jacobian, this will not affect our local qualitative
equations.

\subsection{Quantitative variation of one variable}
We focus 
here on the variation of the concentration of a single chemical
species represented by a component $X_i$ of the vector $X$. Since we
have adopted a {\em static} point of view, we are only interested in
the variation of $X_i$ between two non-degenerated stable steady
states  $X_{eq}^1$ and $X_{eq}^2$ independently of the trajectory of
the dynamical system between the two states.

Let us denote by $\hat{X}_i$ the vector of dimension $n_i$ obtained
by keeping from $X$ all coordinates $j$ that are predecessors of $i$
in the interaction graph. Then, under some additional assumptions
described and discussed in \cite{Radulescu05}, we have the following
result: 

\begin{Theorem}
\label{math:var}
The variation of the concentration
of species $i$ between two non-degenerated steady states $X_{eq}^1$
and $X_{eq}^2$ is given by
\begin{equation}
  X_{eq_i}^1 - X_{eq_i}^2 = \int_S -\left( \DP{F_i}{X_i} \right)^{-1}
  \sum_{k \in \pred(i)} \DP{F_i}{X_k} d X_k 
\label{math:eq:var}
\end{equation}
where $S$ is the segment linking
$\hat{X}^1_{eq_i}$ to $\hat{X}^2_{eq_i}$.
\end{Theorem}

Full proof is given in \cite{Radulescu05}. The above formula is a
quantitative relation between the variation of concentrations and the
derivatives $\DP{F_i}{X_j}$. Now our next move will be to introduce a
qualitative abstraction of this relation.

\subsection{Qualitative equations}
We propose here to study Eq. \ref{math:eq:var} in sign algebra.
By sign algebra, we mean the set $\{\plus,\moins, \indet \}$, where
$\indet$ represents undetermined sign. This set is provided with the
natural  commutative operations:
\[
\begin{array}[c]{llllll}
\plus+\moins = \indet & \plus+\plus = \plus &\moins+\moins = \moins &
\plus\times\moins = \moins & \plus\times\plus = \plus & \moins\times\moins = \plus \\

\indet+\moins = \indet & \indet+\plus = \indet & \indet + \indet = \indet  &
\indet \times \moins = \indet & \indet \times \plus = \indet & \indet
\times \indet = \indet \\
\end{array}
\]

Equality in sign algebra $\approx$ is defined as follows:
\[
\begin{array}{c|c|c|c}
\approx & \plus & \moins & \indet \\
\hline
\plus & T & F & T \\
\hline
\moins & F & T & T \\
\hline
\indet & T & T & T \\
\end{array}
\]

Importantly, qualitative equality is not an equivalence relation,
since it is not
transitive. This implies that computations in qualitative algebra must
be carried with care. At least two major properties should be
emphasized: 
\begin{itemize}
\item if a term of a sum is indeterminate ($\indet$) then the whole
  sum is indeterminate.
\item if one hand of a qualitative equality is indeterminate, then
  the equality is satisfied whatever the value of the other hand is.
\end{itemize}

A \emph{qualitative system} is a set of algebraic equations with
variables in $\{\plus,\moins, \indet \}$. A \emph{solution} of this
system is a valuation of the unknowns which satisfies each equation, 
and such that no variable is instantiated to \indet. This
last requirement is important since otherwise any system would have
trivial solutions (like all variables to \indet).


\begin{Theorem} 
Under the assumptions and notations of Theorem \ref{math:var}, if the
sign of
$\DP{F_i}{X_j}$ is constant, then the following relation holds in sign
algebra: 
\begin{equation}
 s(\Delta X_i) \approx
\sum_{k \in pred(i)}  s(k,i) s(\Delta X_k) \label{math:qual}
\end{equation}
where $s(\Delta X_k)$ denotes the sign of $X_{eq_k}^1 - X_{eq_k}^2$.
\end{Theorem}

By writing Eq. \ref{math:qual} for all nodes in the graph, we
obtain a system of equations on signs of variations, later referred to as
\emph{qualitative system} associated to the interaction graph $G$. This will
be used extensively in the next sections.

\subsection{Link between qualitative and quantitative}
\label{sec:link_qua_quant}

The qualitative system obtained from Eq.\ref{math:qual} is a consequence of
the quantitative relations that result from Theorem \ref{math:var}.  
So the sign function maps a quantitative
variation between two equilibrium points onto a
qualitative solution of Eq.\ref{math:qual}.
The converse is not true in general. For a given solution $S$ of the
qualitative system, there might be no equilibrium change  $\Delta X$
in the differential quantitative model, s.t. each real-valued
component of $\Delta X$ has the sign given by $S$.

However, some 
components of the solution vectors are uniquely determined by the
qualitative system. They take the same sign
value in every solution vector. For such so-called hard components, 
the sign of any quantitative solution (if it exists) is
completely determined by the qualitative system.

We will use the previous  properties to check the coherence between
models and experimental data. By experimental data we mean the sign of
the observed variation in concentration for some nodes. In particular,
if the qualitative system associated to an interaction graph $G$ has
no solution given some experimental observations, then no function $F$
satisfying the sign conditions on the derivatives can describe the 
observed equilibrium shift, meaning that either 
the model is wrong, either some data are corrupted. In the next
section, we introduce a simplified model related to lipid
metabolism, and illustrate the above described formalism.

\section{Toy example: regulation of the synthesis of fatty acids}
\label{sec:we}

\begin{figure*}[hbt]
\begin{center}
\includegraphics[width=12cm]{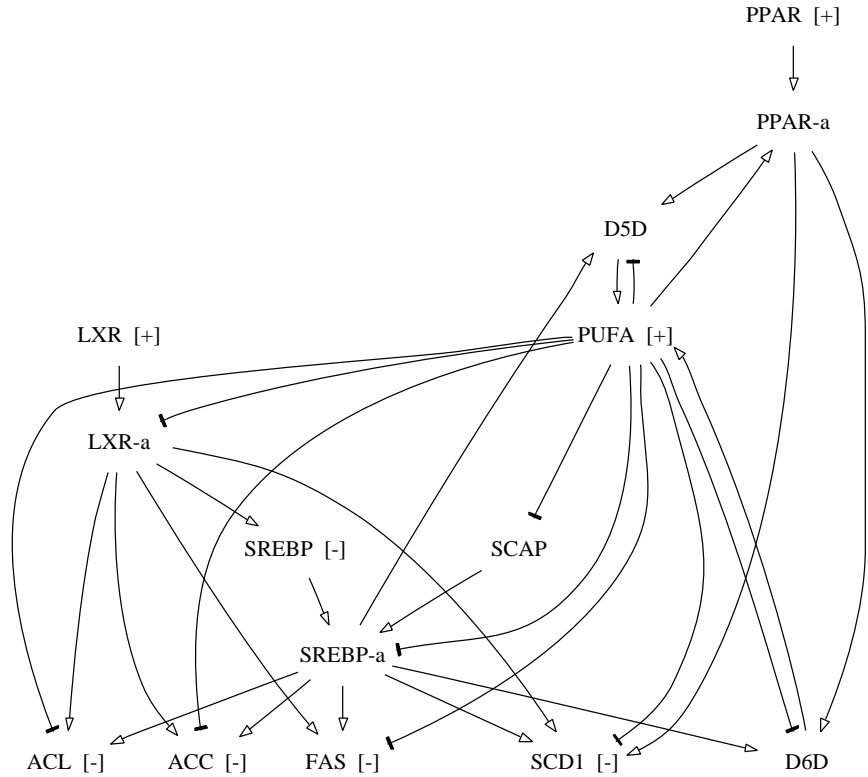}
\caption{Interaction graph for the toy model.
Self-regulation loops on nodes are omitted for sake of clarity.
Observed variations are depicted next to each vertex, when available.}
\label{igraph2}
\end{center}
\end{figure*}

In order to illustrate our approach, we use a toy example describing a
simplified model of genetic 
regulation of fatty acid synthesis in liver. The corresponding
interaction graph is shown in Fig. \ref{igraph2}.

Two ways of production of fatty acids coexist in liver. Saturated
and mono-unsaturated fatty acids are produced from citrates thanks
to a metabolic pathway composed of four enzymes, namely ACL (ATP
citrate liase), ACC (acetyl-Coenzyme A carboxylase), FAS (fatty
acid synthase) and SCD1 (Stearoyl-CoA desaturase 1).
Polyunsaturated fatty acids (PUFA) such as arachidonic acid  and
docosahexaenoic acid are synthesized from essential fatty acids
provided by nutrition;  D5D (Delta-5 Desaturase) and D6D (Delta-6
Desaturase) catalyze the key steps of the synthesis of PUFA.

PUFA  plays pivotal roles in many biological functions; among them, they
 regulate the expression  of genes  that impact on lipid,
 carbohydrate, and protein metabolism. 
The effects of PUFA are mediated either directly through their
specific binding to various nuclear receptors (PPAR$\alpha$ --
peroxisome proliferator activated receptors, LXR$\alpha$ --
Liver-X-Receptor $\alpha$, HNF-4$\alpha$) leading to changes in the
trans-activating activity of these transcription factors; or
indirectly as the result of changes in the abundance of regulatory
transcription factors (SREBP-1c -- sterol regulatory element
binding-protein--, ChREBP, etc.) \cite{Jump}.

\paragraph{Variables in the model}

We consider in our model nuclear receptors PPAR$\alpha$, LXR$\alpha$,
SREBP-1c (denoted by PPAR, LXR, SREBP respectively in the model), as
they are synthesized from the corresponding genes and the
trans-activating active forms of these transcription factors, that is,
LXR-a (denoting a complex LXR$\alpha$:RXR$\alpha$), PPAR-a (denoting a
complex PPAR$\alpha$:RXR$\alpha$) and SREBP-a (denoting the cleaved
form of SREBP-1c. We also consider SCAP -- (SREBP cleavage activating
protein), a key enzyme involved in the cleavage of SREBP-1c, that
interacts with another family of proteins called INSIG (showing the
complexity of molecular mechanism).

We also include in the model ``final'' products, that is, enzymes ACL,
 ACC, FAS, SCD1 (implied in the fatty acid synthesis from citrate),
 D5D, D6D (implied in PUFA synthesis) as well as PUFA themselves.

\paragraph{Interactions in the model}
Relations between the variables are the following. SREBP-a is an
activator of the transcription of ACL, ACC, FAS, 
SCD1, D5D and D6D \cite{Nara,Jump}.
LXR-a is a direct activator of the transcription of SREBP and FAS,
it also indirectly activates  ACL, ACC and SCD1
\cite{StefGus04}. Notice that
these indirect actions are kept in the model  because we don't know
whether they are only SREBP-mediated. 

PUFA activates the formation of PPAR-a from PPAR, and inhibits the formation
of LXR-a from LXR as well as the formation of  SREBP-a (by
inducing the degradation of mRNA and inhibiting the cleavage)
\cite{Jump}. SCAP represents the activators of the formation
of SREBP-a from SREBP, and is inhibited by PUFA.

PPAR directly activates the production of  SCD1, D5D, D6D
\cite{Miller,Tang,Takashi}. The dual regulation of 
SCD1, D5D and D6D by SREBP and PPAR is
paradoxical because SREBP transactivates genes for fatty acid
synthesis in liver, while PPAR induces enzymes for fatty acid
oxidation.  \\ \noindent Hence,  the induction of D5D and D6D gene by PPAR
appears to be
a compensatory response to the  increased PUFA demand caused by
induction of fatty acid oxidation.

\paragraph{Fasting-refeeding protocols}
The  fasting-refeeding
protocols represent a favorable condition for studying
lipogenesis regulation;  we suppose that
during an experimentation, animals (as rodents or chicken) were kept
in a fasted state during several hours. Then, hepatic mRNA of LXR,
SREBP, PPAR, ACL, FAS, ACC and SCD1
are quantified by  DNA microarray analysis. Biochemical measures
also provide the variation of PUFA.

 A compilation of recent
literature on lipogenesis regulation provides hypothetical results of
such  protocols: SREBP, ACL, ACC, FAS and SCD1 decline
in liver during the fasted state \cite{Liang2002}. This is expected
because fasting results in an inhibition of fatty acid synthesis and an
activation of the fatty acid oxidation. For the same reason, PPAR is
increased in order to trigger oxidation. However, Tobin et al
(\cite{Tobin2000}) showed that fasting rats for 24h increased the
hepatic LXR mRNA, although LXR positively regulates fatty acid
synthesis in its activated form. Finally, PUFA levels can be
considered to be increased
in liver following  starvation because of the important lipolysis
from adipose tissue as shown by Lee et al in mice after 72h
fasting (\cite{Lee}).

\paragraph{Qualitative system derived from the graph}
As explained in the previous section, we derive a qualitative system
from the interaction graph shown in Fig. \ref{igraph2}. For ease of
presentation, we denote by \verb|A| the sign of variation for species
A.

{\small
\begin{minipage}{.7\linewidth}
\begin{System}
\begin{verbatim}
(1)  PPAR-a  = PPAR + PUFA
(2)  LXR-a   = -PUFA + LXR
(3)  SREBP   = LXR-a
(4)  SREBP-a = SREBP + SCAP -PUFA
(5)  ACL     = LXR-a + SREBP-a - PUFA
(6)  ACC     = LXR-a + SREBP-a - PUFA
(7)  FAS     = LXR-a + SREBP-a - PUFA
(8)  SCD1    = LXR-a  + SREBP-a - PUFA + PPAR-a
(9)  SCAP    = -PUFA
(10) D5D     = PPAR-a + SREBP-a - PUFA
(11) D6D     = PPAR-a + SREBP-a - PUFA
\end{verbatim}
\label{system2}
\end{System}
\end{minipage}
\begin{minipage}{.3\linewidth}
\begin{Observations}
\begin{verbatim}
  PPAR    = +
  PUFA    = +
  LXR     = +
  SREBP   = - 
  ACL     = -
  ACC     = - 
  FAS     = - 
  SCD1    = -
\end{verbatim}
\end{Observations}
\end{minipage}
}

In the next section, we propose an efficient representation for such
qualitative systems.

\section{Analysis of qualitative equations: a new approach}
\label{sec:eff_qua}

\subsection{Resolution of qualitative systems}
\label{sec:res_qua}

The resolution of (even linear) qualitative systems is a NP-complete
problem (see for instance \cite{Trave,Dormoy88}). One can show this by
reducing the satisfiability problem for a finite set of clauses to the
resolution of a qualitative system
in polynomial time. 

Let us consider a collection $C=\{c_1, \ldots , c_n \}$ of clauses on
a finite set $V$ of variables. Let $\{ \plus ,\moins ,\indet \}$ a sign qualitative
algebra. In order to reduce the satisfiability problem to the
resolution of a qualitative system, let us code $true$ into $+$ and
$false$ into $-$. If $c$ is a clause, let us denote by $\bar{c}$ the
encoding of $c$ in a qualitative algebra formula. The following encoding scheme
provides a polynomial procedure to code a clause into a qualitative
formula. :
$$
\begin{array}{ccc}
\mbox{clause}& & \mbox{sign algebra}\\
\hline
a \in V & \rightarrow & \bar{a}\\
c_1 \vee c_2  & \rightarrow & \bar{c_1} + \bar{c_2}\\
\lnot c & \rightarrow & - \bar{c}
\end{array}
$$

The satisfiability problem for the set of clauses $C$ is then reduced
to finding a solution of the qualitative system:
\[
\{ \bar{c_i} \approx \plus \ /\ i=1, \ldots , n \}
\]
So a NP-complete problem can be reduced to the resolution of
a qualitative system in polynomial time (with respect to the size of
the problem). This shows that solving qualitative systems is a
NP-complete problem.
For example, the only pair of values which are
not solution of $-\bar{a} + \bar{b}\ \approx\ +$ are
$(+,-)$. This corresponds to the only pair $(true, false)$ that does not
satisfy $\lnot a \vee b$. 

Several heuristics were proposed for the resolution of qualitative
systems. For linear systems, set of rules have been
designed \cite{Dormoy88}. This set is complete: it allows to find every
solution. It is also sound: every solution found by applying these
rules is correct. The rules  are based on an adaptation of Gaussian
elimination. However only heuristics
exist for choosing the equation and the rule to apply on it. In case
of a dead-end, when no more rule can apply, it is necessary to backtrack
to the last decision made. As a result programs implementing
qualitative resolution are not very
efficient in general and only problems of small size can be resolved
in reasonable time. For that reason we propose an alternate way to
solve qualitative systems (linear or not).

\subsection{Qualitative equation coding}
\label{sec:eq_coding}

Our method is based on a coding of qualitative equations as
algebraic equations over Galois fields ${\mathbb Z}/p{\mathbb Z}$
where $p$ is a prime number greater than 2. The elements of these 
fields are the classes
modulo $p$ of the integers. If $\bar{x}$ denotes the class of the
integer $x$ modulo $p$, a sum and a product are defined on ${\mathbb
Z}/p{\mathbb Z}$ as follows:
$$ 
\bar{x} + \bar{y}  =  \overline{x+y} \qquad \bar{x} \times
\bar{y}  =  \overline{x \times y}
$$

Galois fields have two basic properties which we use extensively:
\begin{itemize}
\item Every function $f:({\mathbb Z}/p{\mathbb Z})^n \rightarrow
      {\mathbb Z}/p{\mathbb Z}$ with  $n$  
  arguments  ${\mathbb Z}/p{\mathbb Z}$ is a polynomial function
\item if $\oplus$ denotes the operation $f \oplus g = f^{(p-1)} +
      g^{(p-1)}$, then every equation system  $p_1(X)=0, \ldots ,
      p_k(X)=0$ has the same solutions than the unique equation $p_1\oplus p_2 \oplus
  \ldots \oplus p_k (X) = 0$.
\end{itemize}

The following table specifies how the sign algebra
$\{\plus,\moins,\indet\}$ is mapped onto the Galois field with three
elements ${\mathbb Z}/3{\mathbb Z}$ is used for that coding. 
$$
\begin{array}{ccc}
\mbox{sign algebra}& & \mathbb{Z}/3{\mathbb Z}\\
\hline
\plus  & \rightarrow &  1\\
\moins & \rightarrow & -1\\
\indet & \rightarrow &  0\\
\end{array}
\qquad \qquad \qquad
\begin{array}{ccc}
\mbox{sign algebra}& & \mathbb{Z}/3{\mathbb Z}\\
\hline
e_1 + e_2 & \rightarrow & - \overline{e_1}.\overline{e_2}.(\overline{e_1} + \overline{e_2})\\
e_1 \times e_2 & \rightarrow & \overline{e_1}.\overline{e_2}\\
e_1 \approx e_2 & \rightarrow & \overline{e_1}.\overline{e_2}.(\overline{e_1} - \overline{e_2})
\end{array}
$$

Finally a qualitative system $\{ e_1,\dots,e_n \}$ is coded as the
polynomial $\overline{e_1} \oplus \dots \oplus \overline{e_n}$.
A similar coding for the qualitative
algebra $\{\plus,\moins,\zero,\indet\}$ uses the Galois field 
${\mathbb Z}/5{\mathbb Z}$ and will not be presented here.

With this coding, every qualitative system has a solution if and only
if the corresponding polynomial has a solution without null
component. Null solutions are excluded since $\indet$ solutions are
excluded for qualitative systems. In general we will have to add
polynomial equations $X^2 = 1$ to insure this.

\subsection{An efficient representation of polynomial functions}
\label{sec:rep_func}

Recall that our purpose is to efficiently solve a NP-complete
problem. There is no hope to find a representation of polynomial
functions allowing to solve polynomial systems of equations in
polynomial time. The coding of a qualitative system as a
polynomial equation is obviously polynomial in the size of the system
(number of variables plus number of equations). So finding the
solution of a polynomial system of equations is itself a NP-complete
problem. It is more or less the SAT problem.

Nevertheless, there exists a representation of polynomial functions on
Galois fields which gives, in practice, good performances for
polynomials with hundreds of variables. This kind of representation
was first used for logical functions which may be considered as
polynomial functions over the field ${\mathbb Z}/2{\mathbb Z}$. This
representation is known as BDD (Binary Decision Diagrams) and is
widely used in checking logical circuits \cite{BDD} and in model checkers
as nu-SMV \cite{SMV}.

We present here this representation for the field ${\mathbb
Z}/3{\mathbb Z}$. Generalizations to other Galois fields 
could be treated as well. The starting point is a generalization of Shannon
decomposition for logical functions:
\[
p(X_1,X) = (1-X_1^2)p_{[X_1=0]}(X)+X_1(-X_1-X_1^2)p_{[X_1=1]}(X) +
X_1(X_1-X_1^2)p_{[X_1=2]}(X) 
\]
where $p$ is a polynomial function with $n$ variables. This
decomposition leads to a tree representation of the polynomial
function: the variable $X_1$ is the root and has three children. Each
of these is obtained by instantiating $X_1$ to -1, 0 or 1 in
$p(X_1,X)$. This representation is exponential ($3^n$) as each non
constant node has 3 children. It also depends on a chosen order on
the variables.

Then a key observation (see \cite{BDD}), is that several
subtrees are identical. They have the same variable as root variable
and isomorphic children. If we decide to represent only once each type
of tree, then the tree representation is transformed into  a direct
acyclic graph. With this representation there is no more redundancy
among subtrees. The result may be a dramatic decrease in the size of
the representation of a polynomial function.

\begin{figure*}[hbt]
\begin{center}
\includegraphics[width=12cm]{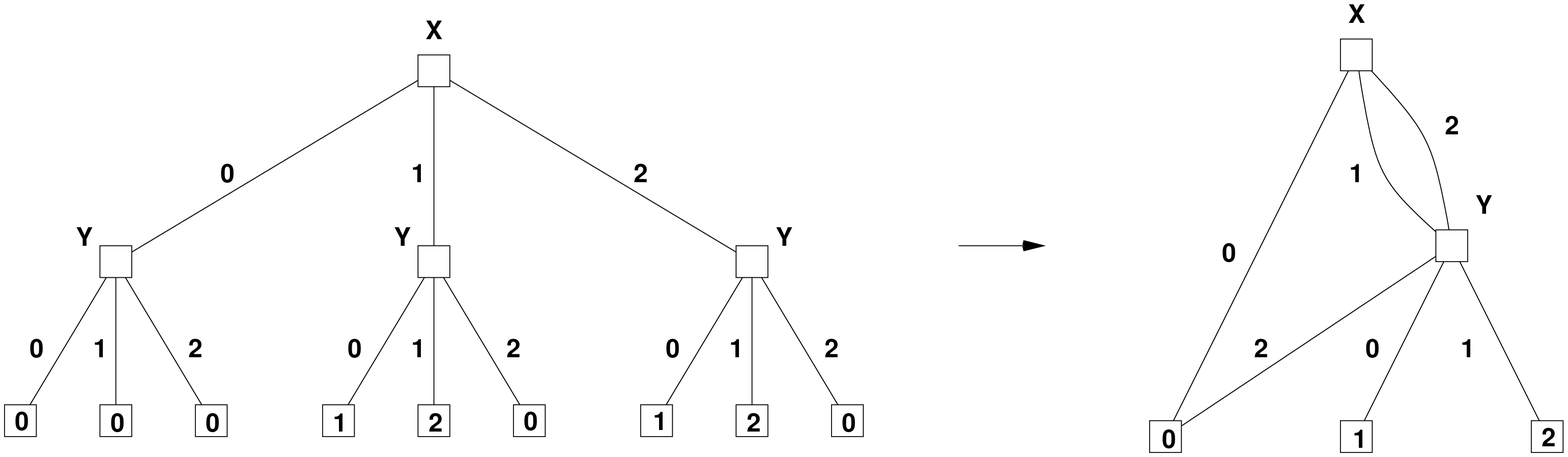}
\caption{From tree representation to direct acyclic graph for $X^2
(Y+1)$. The tree has 13 nodes while the DAG representing the same
function has 5 nodes.}\label{itree-dag}
\end{center}
\end{figure*}

A property of the Shannon like decomposition is that many
operations on polynomial functions are recursive with respect to this
decomposition. More precisely let
\[
p^i(X_1,X) = (1-X_1^2)p^i_0(X)+X_1(-X_1-X_1^2)p^i_1(X) +
X_1(X_1-X_1^2)p^i_2(X) 
\]
$i=1,2$ be two polynomial functions with $p_{\alpha}(X) = p_{[X_1 =
\alpha]}(X)\ ,\ \alpha = 0,1,2$. Then for binary operations $\Delta$
on polynomial  functions,
\[
p^1 \Delta p^2 = (1-X_1^2) (p^1_0 \Delta p^2_0 ) + X_1(-X_1-X_1^2)
(p^1_1 \Delta p^2_1 )+  X_1(X_1-X_1^2) ( p^1_2 \Delta p^2_2 )
\] 
This kind of recursive formula leads to an exponential
complexity of any computation. Again, it is possible to take advantage
of the redundancy by using a cache to remember each operation. This
technique is known as
memoisation in formal calculus. A  40\% cache hit rate is commonly observed. 

More complex operations on polynomial functions are also implemented
with a recursive scheme and memoisation. Let us just mention
quantifier elimination as among the most useful for our
purpose. 

This representation of polynomial functions on Galois fields has
also several drawbacks:
\begin{itemize}
\item the memory size heavily depends on the order of variables. The
      libraries implementing formal computations always have reordering
      algorithms. 
    \item for each order, there exists polynomial functions which are
          exponential in memory size.
\end{itemize}

Nevertheless, in practice, this representation has proved to be very
efficient for polynomial functions with several hundred of
variables. The computations performed on our toy model and on another 
real size one
used a program named SIGALI which is devoted to polynomial functions on
${\mathbb Z}/3{\mathbb Z}$ representation. Several algorithms were
added to this program in order to answer questions of biological
interest. 

\section{Qualitative models and experimental data}
\label{sec:qua_exp}
In this section, we show how to compute some properties of a
qualitative system, and eventually get some insights on the biological
model it represents. The algorithms we derive heavily rely on the
representation introduced above. Hence, not only they can deal in
practice with computationally hard problems efficiently, but also they
are expressed in a rather simple and generic fashion.

Let $M$ be a qualitative model represented by its associated
interaction graph $G(V,E)$. Recall that $V$ is the set of
variables. Let $V_O$ be the set of observed variables, and 
$o_i \in \{ \plus,\moins \}$ for $i \in V_O$ the experimental
observations. As explained in the previous section, the 
qualitative system derived from $M$ can be coded as a polynomial
function $P_M(X_1,\ldots,X_n)$.  Roots of $P_M$ correspond to
solutions of the qualitative system. 

\subsection{Satisfiability of the qualitative system}
\label{sec:data_agree}

A property of the coding described above, is that the system has no
solution iff $P_M$ is equal to the constant polynomial
$1$. Alternatively if $P_M=0$, the qualitative equations do not
constraint the variables at all.
 
Now if some observations $o_i$ for $i \in V_O$ are available, checking
their consistency with the model $M$ boils down to instantiating 
$X_i = o_i$  in $P_M(X_1,\ldots,X_n)$, for all $i \in V_O$, and
testing whether the resulting polynomial is different from 1.

We computed the polynomial $P_L$ associated to our toy example (see section
\ref{sec:we}) and it has roots. Recall that it does not guarantee the
existence of some (quantitative) differential model conforming to the
interaction graph depicted in Fig. \ref{igraph2}. Satisfiability of
the qualitative system is only a necessary condition for the model to
be correct.



The polynomial obtained by instantiating observations into $P_L$
is different from 1, meaning that our model does not contradict
generally observed variations during fasting.

Large size models might advantageously be reduced
using standard graph techniques.
First we look for connected components in the interaction graph. A
graph with several
connected components represents a coherent qualitative model iff 
each component is coherent. 
Second, a node without successor except itself appears only
in its associated equation. If this node is not observed, its
associated qualitative equation adds no constraint on the other
nodes. So, at least for satisfiability checking, this node can be
suppressed and its qualitative equation removed from the system. This
procedure is applied iteratively, until no node can be deleted.
The resulting graph leads to a new qualitative system which is 
satisfiable iff the initial system is satisfiable.

\subsection{Correcting data or model}

If the qualitative system, given some experimental observations, is
found to have no solution, it is of interest to propose some
correction of the data and/or the model. By correction, we mean
inverting the sign of an observed variable or the sign of an edge of
the interaction graph. In
the general case, there are several possibilities to make the system
satisfiable, and we need some criterion to choose among them. We
applied a parsimony principle: a correction of the data should imply a
minimal number of sign inversions.

In the following, we show how to compute all minimal corrections for the
data. Given $(o_i)_{i \in V_O}$ a vector of experimental observations
which is not compatible with the model, we compute all 
$(o'_i)_{i \in V_O}$ vectors which are compatible with the data and
such that the Hamming distance between $o$ and $o'$ is minimal. By
Hamming distance, we mean the number of differences between $o$ and
$o'$. The set of such $o'$ vectors might be very large; but again, by 
encoding it as the set of roots of a polynomial function,  we obtain a 
compact representation.

This procedure can be extended in a straightforward manner to
corrections of edges sign in the interaction graph. This is done by
considering these signs as variables of the model. For ease of
presentation, we only detail data correction.

\begin{algorithm}
{
  \small
  \dontprintsemicolon
  \KwIn{\\
    \qquad $P$, a polynomial function on variables $V$ \\
    \qquad $i \in V$ 
  }
  \KwOut{\\
    \qquad $C$, a polynomial function encoding all minimal corrections\\
    \qquad $d$, minimal number of corrections
  }
  \BlankLine
  \eIf{$P$ is constant}{
    \eIf{$P$ = 0}
   {\KwResult{$C=0$, $d=0$}}
   {\KwResult{$C=1$, $d= \infty$}}
  }{
    let $P_0$, $P_1$, $P_2$ be the Shannon decomposition of $P$ with respect to
    variable $X_i$,\\
    and $(C_j,d_j)$ the result obtained by recursively applying the
    algorithm on $P_j$ and $i+1$ for $j \in \{0,1,2\}$
    \BlankLine
    let $d'_j = 
    \left\{
    \begin{array}{ll}
      d_j + 1 & \mbox{if } i \in V_O \mbox{ and } o_i \neq j\\
      d_j     & \mbox{otherwise}
    \end{array}
    \right.$
     and $C'_j = 
    \left\{
    \begin{array}{ll}
      (X_i - j) \oplus C_j & \mbox{if } i \in V_O \\
      C_j     & \mbox{otherwise}
    \end{array}
    \right.$\\
     \BlankLine
    
     \KwResult{$d = \min\ d'_j$, $C = \displaystyle\prod_{j,\ d'_j = d} C'_j$}
  }
}
\caption{Algorithm for experimental data correction.}
\label{algo:hamming}
\end{algorithm}

Let us illustrate this algorithm on our toy example: during fasting
experiments, synthesis of fatty acids tends to be inhibited, while
oxidation, which produces ATP, is activated. In particular ACC, ACL,
FAS and SCD1 are implied in the same pathway to produce saturated and
monounsaturated fatty acids. Expectedly, they are known to decline
together at fasting. Suppose we introduce some wrong observation, say
for instance an increase of ACL, while keeping all other
observations given above. The polynomial obtained from $P_L$ including
these new observations is equal to 1, and hence has no solution.
Applying algorithm \ref{algo:hamming}, we recover this error. Now if
we wrongly change two values, say ACL and ACC to 1, the algorithm
proposes a different correction, namely to change the observed value
of SREBP to 1, which is more parsimonious.

\subsection{Experiment design}
\label{sec:exp_des}

It is often the case that not all variables in the system under study
can be observed. Biochemical measurements of metabolites can be costly
and/or time consuming. By experiment design, we mean here the choice
of the variables to observe so that an experiment might be
informative. 

Let $P_M(X_O,X_U)$ be the polynomial function coding for the
qualitative system $M$. $X_O$ (resp. $X_U$) denotes the state vector
of observed (resp. unobserved) variables. The polynomial function
representing the admissible values of the observed variables is
obtained by elimination of the quantifier in 
$\exists X_U \  P_M(X_O,X_U)$. Let $P_M^O(X_O)$ denote the  resulting
polynomial function.

For some choice of observed variables, it might well be that $P_M^O$
is null, which basically means that the experiment is totally useless.
Remark that no improvement can be found by taking a subset of
$X_O$  The solution is either to add new observed variables or to
chose a completely different set of observed variables.

In order to assess the relevance of a given experiment (namely of a
given observed subset), we suggest to compute the following ratio:
number of consistent valuations for observed variables versus the
total number
of valuations of observed variables. A very stringent experiment
has a low ratio. An experiment having a ratio value of one is useless.

Again this computation is carried out in a recursive fashion. Let $P$
be a polynomial function representing the
set of admissible observed values. Let $Rat(p)$ the percentage of
solutions of $P(X)=0$ in the
space $({\mathbb Z}/p{\mathbb Z})^n $, where $n$ is the number of
variables $X$. If $P$ is constant then $Rat(P)=1$
(resp. $Rat(P)=0$) if $P=0$ (resp. $P \neq 0$). Else, let $P_1$,
$P_2$, $P_3$ be a Shannon like decomposition of $P(X)$
with respect to some variable of $P$. Then it is easy to prove:
\[
Rat(P)\ =\ (Rat(P_0)+Rat(P_1)+Rat(P_2))/3
\]

The relevance of this approach was assessed on our toy example: for
each subset $O$ of variables in the model, containing at most four
variables, we computed $Rat(P_L^O)$. Expectedly, the lowest ratios
(i.e. the most stringent experiments) were achieved observing four
variables: either \{SCAP, PUFA, PPAR-a, PPAR\}, or
\{SREBP, SCAP, PUFA, LXR-a\}, or \{SREBP, PPAR-a, PPAR, LXR-a\}.

Interestingly, the procedure captures what might be though of as
control variables, like PUFA/SCAP, SREBP/LXR-a and PPAR/PPAR-a.
The first two pairs control the activation of fatty acids synthesis;
the third one controls fatty acid oxidation.

Indeed one can go even further: if we isolate some kind of control
variables, we are naturally interested in knowing how they constrain
other variables. Achieving this amounts to computing the set of
variables which value is constant for all solutions of the system (the
so called hard components). Recall that
these hard components of qualitative solutions are also important with
respect to the hypothetical differential model which is abstracted in
the qualitative one. Indeed, all solutions of the quantitative equation for
equilibrium change have the same sign pattern on the hard components. 
Algorithm \ref{algo:hardcomponents} describes a recursive procedure
which finds the set of hard components, together with their value. 

\begin{algorithm}
{
  \small
  \dontprintsemicolon
  \KwIn{$P$, a polynomial function on variables $V$}
  \KwOut{\\
    \qquad the set $W \subset V \times \{0,1,2\}$ of hard components,
    together with their values\\
    \qquad a boolean $b$ which is true if $P$ has at least one root
  }
  \BlankLine
  \eIf{$P$ is constant}{
    \eIf{$P$ = 0}
   {\Return{$(\emptyset,\mbox{true})$}}
   {\Return{$(\emptyset,\mbox{false})$}}
  }{
    let $P_0$, $P_1$, $P_2$ be the Shannon decomposition of $P$ with respect to
    variable $X_i$,\\
    and $(W_j,b_j)$ the result obtained by recursively applying the
    algorithm on $P_j$ for $j \in \{0,1,2\}$
    \BlankLine
    let $W = \{ (v,v') | v \in V,\ v' \in \{0,1,2\},\ \forall j\ b_j \Rightarrow (v,v') \in W_j \}$\\
    \If{there exists a unique $j_0$ s.t. $b_{j_0}$ is true}
       {add $(i,j_0)$ to $W$}
    \BlankLine
    \Return{$(W,b_0 \vee b_1 \vee b_2)$}
  }
}
\caption{Determination of hard components}
\label{algo:hardcomponents}
\end{algorithm}

Let us set some of our previously found control variables of the toy example,
to a given value, say PUFA to 1, and LXR to -1. Then applying the
algorithm \ref{algo:hardcomponents}, the corresponding polynomial has the
following hard components: 
\begin{verbatim}
ACL     = -1             FAS   = -1
ACC     = -1             LXR-a = -1
SCAP    = -1             SREBP = -1
SREBP-a = -1             PPAR  = -1
PPAR-a  = -1
\end{verbatim}
which expectedly corresponds to the inhibition of fatty acids
synthesis.

\subsection{Real size system}
 We have used our new technique to check the consistency of a
database of molecular interactions involved in the genetic regulation of fatty
acid synthesis.
In the database, interactions were classified as behavioral or biochemical. 
\begin{itemize}
\item a behavioral interaction describes the effects of a variation of
      a product concentration. It is either direct or indirect (unknown
      mechanism).
    \item a biochemical interaction may be a gene transcription, a
          reaction catalyzed by an enzyme ... Such molecular
          interactions can be found in existing databases. They need
          a behavioral interpretation.
\end{itemize}
 All the behavioral interactions were manually extracted from a 
selection of scientific papers. Biochemical interactions were
extracted from  public databases available on the Web 
(Bind~\cite{Bind}, IntAct~\cite{IntAct}, Amaze~\cite{Amaze},
KEGG~\cite{Kegg} or TransPath~\cite{TransPath}).
A biochemical interaction may be linked to a behavioral interpretation
in the database.

The database is used to generate the interaction graph. While
behavioral interactions directly correspond to edges in the graph, 
biochemical interactions are given a simplified
interpretation. Roughly, any increase of a reaction input induces an
increase of the outputs.

The interaction graph which is built from the database contains more than 600
vertices and more than 1400 edges. It is clear that even though, the
obtained graph is not a comprehensive model of genetic
regulation of fatty acid synthesis in liver. Anyway our aim is to see
how far this model can account for experimental observations, and
propose some corrections when it cannot.

We used our technique to check the coherence of the whole model. After
reducing the graph with standard graph techniques as described in
section \ref{sec:data_agree}, we found that the model was incoherent. The reduced
graph has about 150 nodes. We
developed a heuristic to isolate minimal incoherent sub-systems. It
turned out that all the contradictions we detected resulted from arguable
interpretations of the literature.

\section{Conclusion}
\label{sec:conclusion}

In this paper we proposed a qualitative approach for the analysis
of large biological systems. We rely on a framework more thoroughly
described in \cite{Biosystems}, which is meant to model the comparison
between two experimental conditions as a steady state shift. This
approach fits well with
state of the art biological measurement techniques, which provide
rather noisy data for a large amount of targets. It is also 
well suited to the use of biological knowledge, which is most of the time 
descriptive and qualitative. 

This qualitative approach is all the more attractive that we can rely
on new analysis methods for qualitative systems. This new technique is
also introduced in this paper and is original in
qualitative modeling. It relies on a representation of qualitative
constraints by decision diagrams. Not only this has a major impact on
the scalability of qualitative reasoning, but it also permits to
derive many algorithms in a quite generic fashion.

We plan to validate our approach on pathways which are published for
yeast and \emph{E.Coli}. Not only this pathways are of significant size but
microarray data for this species are publicly available. Concerning
the scalability of the methods, qualitative systems with up to 200
variables are handled within a few minutes.

On the theoretical side, we study applications of our algebraic
techniques to network reconstruction, as proposed in \cite{Wagner04}.
The problem is to infer direct actions between products, based on
large scale perturbation data, in order to obtain the most
parsimonious interaction graph. Our approach could lead to a
reformulation of this problem in terms of polynomial operations. 
Indeed, finding a minimal regulation network from a minimal polynomial
representation has already been described in
\cite{Laubenbacher04}, though it was applied to a rather different
type of network. A similar approach tailored to the framework
described in this paper could eventually lead to original and
practical algorithms for network reconstruction.

\paragraph{Acknowledgment} This research was supported by ACI IMPBio, a French Ministry for Research program on interdisciplinarity.

\bibliographystyle{plain}
\bibliography{eccs}
\end{document}